\documentclass[journal]{IJIT}
\usepackage{graphics,graphicx}

\begin{document}

\title{openPC : a toolkit for public cluster with full ownership}

\author{Z. Akbar, I. Firmansyah, B. Hermanto, L.T. Handoko
        \thanks{All authors belong to the Group for Theoretical and
Computational Physics, Research Center for Physics, Indonesian Institute of
Sciences, Kompleks Puspiptek Serpong, Tangerang 15310, Indonesia, contact email
: gftk@mail.lipi.go.id.

Z. Akbar is also with the Group for Bioinformatics and Information
Mining, Department of Computer and Information Science, University of Konstanz,
Box D188, D-78457 Konstanz, Germany.

I. Firmansyah is also with the Center for Environmental Remote Sensing,
Chiba University, 1-33 Yayoi-cho, Inage-ku, Chiba, Japan.}}

\maketitle
\thispagestyle{empty}

\begin{abstract}
The openPC is a set of open source tools that realizes a parallel machine and
distributed computing environment divisible into several independent blocks of
nodes, and each of them is remotely but fully in any means accessible for  users
with a full ownership policy. The openPC components address fundamental issues
relating to security, resource access, resource allocation, compatibilities with
heterogeneous middlewares, user-friendly and integrated web-based interfaces,
hardware control and monitoring systems. These components have been deployed
successfully to the LIPI Public Cluster which is open for public use. In this
paper, the unique characteristics of openPC due to its rare requirements are
introduced, its components and a brief performance analysis are
discussed.
\end{abstract}


\IEEEpeerreviewmaketitle

\section{Introduction}

The popularization of high performance computing (HPC) leads to great demands on
highly skilled human resources in the field. Although, in recent years there are
many emulators providing a practical platform for parallel programming on a
single PC, comprehensive skills on parallel programming can in most cases be
obtained by the users having comprehensive distributed environments with full
ownership and access. Moreover, the parallel programming highly depend on the
problems under consideration and the architectures of distributed computing
environment where the codes are going to be executed on. Though building a small
scale HPC is getting easier and cheaper in recent days, yet it requires
unaffordable cost and expertise for most people or small research groups. In
particular, a 'comprehensive and advanced' HPC environment actually needs more 
non-trivial efforts to realize. The comprehensive and advanced HPC  environment
here means the advanced environment deployed in the system (middleware, resource
allocation management, etc), and is not just the great number of nodes.

Therefore enabling as wide as possible access to an advanced HPC machine is
getting an important issue in, especially educating young generation with
comprehensive knowledge in the field. As a key solution to overcome this issue
one of us \cite{lpc}, has developed an alternative architecture for parallel
environment available for public use called as LIPI (Lembaga Ilmu Pengetahuan
Indonesia - Indonesian Institute of Sciences) Public Cluster (LPC). The initial
prototype has succeeded in enabling full access for remote users to the whole
aspects of HPC over an integrated web interface  \cite{hakcipta}. In LPC the web
interface plays an important role to :
\begin{itemize}
\item Provide a user friendly interface concerning that most users are at the
beginner level. Because LPC is originally intended for educational purposes,
i.e. providing a training field to learn all aspects of distributed programming
and computing environment.
\item Keep the whole system secured from any potential attacks since the LPC is open for
anonymous users. Everyone can start using the system once their online
registration was approved. Though this simple procedure is definitely reducing
the obstacles for users, instead the administrator has to pay special attention to secure
the system. Hence, the main issue in this matter is how to replace the usual
\texttt{ssh} based commands with the web based ones, while keeping the same
level of degree of freedom.
\end{itemize}
During its first phase of implementation, the LPC enjoyed rousing welcome from
the communities. We should remark that although the LPC was intended for
educational purposes, the facility has also been used by some small research
groups to perform more serious scientific tasks. 

In the second phase of development, we have improved all existing components of
LPC and also added some new components to meet recent requirements to be a real
HPC machine. Those  involve the basic architectures \cite{rict}, the algorithm
of multi block approach \cite{iceei}, the resource allocation \cite{icici2} and
the real-time control and monitoring systems \cite{icici1,qir}. Finally we have
recently bundled those components into an integrated toolkit, namely the Open
Public Cluster (openPC) and release it for public use under GNU Public License
\cite{openpc}.

In the present paper, we are introducing the current status of openPC, its
concepts and detailed architectures. We also present the current implementation
at the LPC which continues providing its services in Indonesia and surrounding
regions. The paper is organized as follows. First the basic concept of LPC is introduced. It is followed by the global architecture and the detail of queue and resources management schemes. Before giving the conclusion the implementation of current system at LPC is explained.

\section{CONCEPTS}

\begin{figure*}
\centerline{\includegraphics[width=14cm]{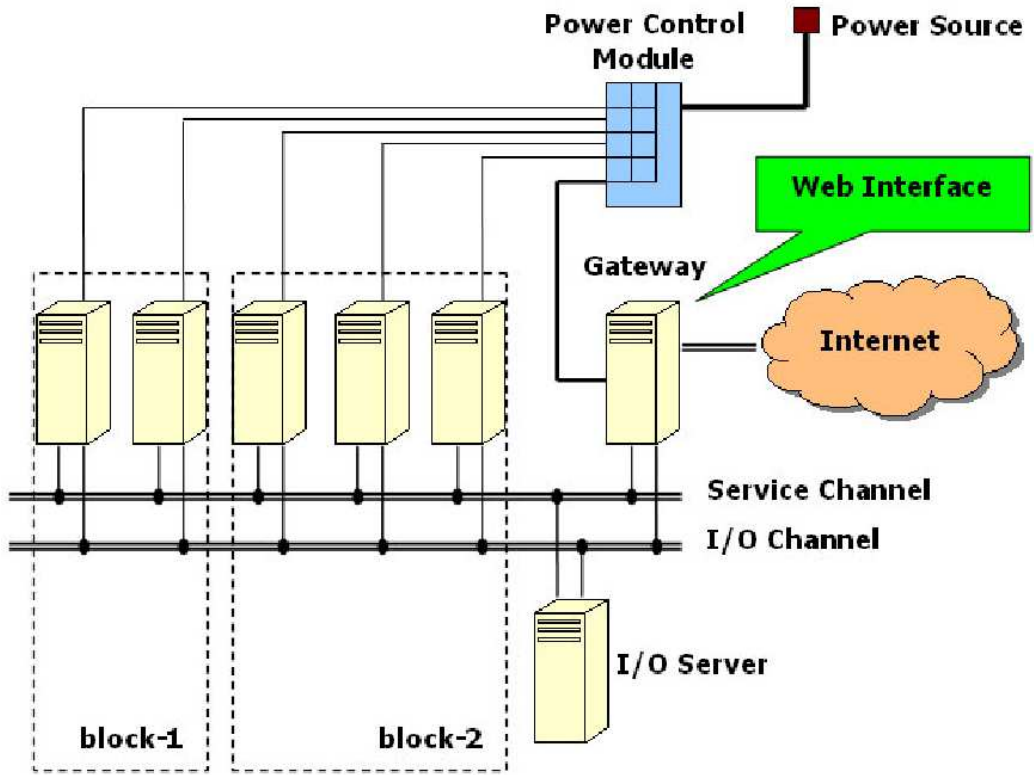}}
\caption{The concept of public cluster with full ownership on each block of
nodes using multi-block approach.}
\label{fig:diagram}
\end{figure*}

As already mentioned briefly in the preceding section, the concept of public
cluster is motivated mainly by a need to provide a public infrastructure for
HPC. One of us has built a successful prototype of LPC with the following
purposes \cite{lpc} :
\begin{itemize}
\item Providing a practical but real training field for teachers and students to
explore any aspects of HPC, that is not only developing the parallel programming
itself.
\item Improving the utilization of the existing HPC facilities. Since in some
HPC centers the infrastructures tend to idle away due to limited users.
\end{itemize}
Clearly, these purposes are more relevant for some developing countries with
limited HPC resources. However, the issues are, especially the second one is,
actually relevant to any other developed regions as well. The limitations are in
general caused by tight regulations due to mostly security concerns.

The first purpose above requires ``full ownership`` by users on the cluster
being used. It cannot be replaced by conventional clusters where the users just
put their parallel jobs on the queues and let the machines do the rest. The full
ownership here is defined as the user is provided with enough privileges to : 
\begin{itemize}
\item Virtually, but fully own a block of nodes within a certain period without
any job  from the others in the same block.
\item Choose any parallel environment, i.e. the middleware for job
distributions, etc.
\item Allocate the resources according to the user preferences within an
assigned block.
\end{itemize}
Just to mention, in the near future the block owned by a user is able to be
connected to any global grid relevant to the user's project.

These concepts  enable users to explore and perform any types of
computational works  on an HPC machines. This is the main reason we have argued
that the system is suitable for educational purposes. Of course, the advanced
users could also benefit from the system, since they can use it to perform any
distributed computations with much more degrees of freedom. In contrast to any
existing HPC machines, the public cluster is becoming a kind of really general
purpose HPC machine. This would also improve the utilization and be good
justification to obtain public funding to initiate and maintain an HPC center as
well.

In order to realize the concepts, we have proposed a simple solution using the
so-called multi-block approach \cite{iceei}. As shown in Fig. \ref{fig:diagram},
the public cluster, or just a set of parallel machine, is divided into several
inhomogeneous blocks. Each of them consists of various number of nodes and forms
a subset of resources. Utilizing the microcontroller-based control and
monitoring systems, each node is controlled and monitored independently over web
through the gateway. Throughout the paper we call the subset of nodes as a
block.

We should remark here that according to the above mentioned concepts, openPC has
completely different natures with another existing web-based clustering toolkits
to enable ''remote access over web`` like OSCAR \cite{oscar}, or some grid
portals for interfacing large scale distributed systems over web \cite{ganglia,
corba,hpc2n, gird, gridspace}. On the other hand, it does also not belong to the
same category as Globus Toolkit (GT) which is intended to ''interface`` large
scale grid computing \cite{gt}. We also recognize some partial web-interfaces
for clusters developed by several groups as done in \cite{sce, shibboleth,
webmpi}. Though the architecture of openPC is like a miniature of grid, since
the blocks are connected each other as a single conventional parallel machine
the physical topology is quite different. Moreover, openPC has a capability to
control and monitor each single node that is irrelevant in tools like GT.

\section{ARCHITECTURE}

The openPC is a toolkit set dedicated to realize the concepts of public cluster.
It consists of several components, ranging from hardware control to web
interface. The toolkit has been developed with main purposes :
\begin{itemize}
\item As a single interface between users and the blocks assigned for each user
with full ownership policy. This includes integrating all components and making
them accessible to remote users over a user-friendly web interface.
\item Providing as high as possible degree of freedom to users, while on the
other hand keeping the whole system totally secured. The main concern is
especially to keep the user and their jobs stay on the assigned blocks.
\end{itemize}
Under the full-ownership policy, a user is provided exclusively an HPC system
without worrying about potential interferences with the others. All complexities
are concealed behind the screen, although the users are still able to configure
their own block as if an independent HPC machine like choosing the middleware
for distributed environment and so on.

When a user submits a particular command through the web interface, the openPC
performs several internal processes :
\begin{enumerate}
\item Recognizing whether the command is related to : the distributed computing,
i.e. Portable Batch System (PBS), the operating system (OS), or the hardwares,
i.e. \texttt{port com}. This selection procedure is crucial to determine how to
treat and the forwarded target of each command.
\item If the command is a \texttt{port com} type, the openPC calls the library
to execute further an  appropriate command to the hardware modules (control and
monitoring systems). In contrary, the PBS or OS type commands are passed to the
I/O / master node through a newly developed layer called public cluster module
as secure shell (\texttt{ssh}) wrapped commands.
\item On the other hand, each node within the block has two daemons running
simultaneously, that is MoM (Machine Oriented Mini-server) and MPD. The PBS MoM
is responsible for job management, while the \texttt{mpd} is the computation
daemon to enable message passing among the nodes. 
\end{enumerate}
With these procedures and the hardware configuration depicted in Fig.
\ref{fig:diagram}, we can move forward to discuss the details.

\subsection{Mapping a queue to a subset of resources}

As briefly mentioned before, one of the main concerns is how to keep a user's
jobs are sent to the designated nodes within the user's block. This can be
achieved by employing a mapping procedure. In openPC, we have borrowed the
Terascale Open-source Resource and QUEue manager (TORQUE) and Maui Scheduler to
realize that requirement \cite{torque, maui}. TORQUE is actually an extension of
PBS or definitely OpenPBS \cite{openpbs}. On the other hand, Maui is a "policy
engine" to control "when, where, how" the available resources like processor,
memory and disk space are allocated to particular jobs in the queue. Maui does
not only provide an automated mechanism to optimize the available resources, but
also to monitor the performances and analyze the problems occurred during the
running periods.

Utilizing the TORQUE and Maui, the mapping procedure in openPC is done as
follows :
\begin{enumerate}
\item Creating the host access list :\\
The host access list is configured and activated in each queue such that only
those nodes  can recognize the queue. This will ensure that the queue is sending
the jobs to the assigned nodes of a particular block owned by the user.
\item Creating the user access list :\\
This access list limits the users who are allowed to submit jobs to particular
queues. This method is used to avoid, either intentionally or unintentionally,
submission to another queues.
\end{enumerate}

Once a block has been created in a public cluster, it will remain with its
initial nodes and user within the approved period of usage. This is suitable for
the main purpose under consideration, that is providing a training field for
learning HPC environment. The  exclusive blocks allow the users to perform
experimental jobs with high degree of freedom, or to explore any aspects of HPC,
without any interferences among them. Moreover, in the current openPC each block
has its own master node to improve the user degree of freedom, while reducing
the load on I/O server. This allows the users to choose the desired parallel
environment etc for their blocks. Anyway, it has been shown that up to a certain
extent the multi blocks approach with single master node is still reliable
\cite{icts}.

Through these algorithms, we can define a queue which represents a closed subset
of nodes  owned by a single user without being interfered by another ones. This
simple mechanism is the underlying idea to provide those blocks to public with
full ownership. Therefore a block is a subset of nodes defined in a single
queue.

\subsection{Resources management scheme}

A resource manager always attempts to choose the best combination of nodes to
perform the submitted jobs automatically. In the conventional cluster, the users
should determine how to distribute the jobs over the nodes manually by fixing
certain parameters, or choosing among available queues. On the other hand, the
administrator set up some queues for certain purposes which can be chosen by
the users when submitting the jobs.

In contrary with this in the LPC, however we can not deploy the same approaches
since the users (and also their jobs) are anonymous, while they basically own
the blocks fully. Since in the LPC a queue is created once a block has been
activated, the queue is exclusively limited to a particular user and consists of
certain nodes assigned to the user. Hence the users have full privileges to
allocate the jobs and manage the nodes within their blocks. Using TORQUE and
Maui. resource management within a block is performed in the same way as
conventional clusters.

\section{IMPLEMENTATION}

\begin{figure*}
\centerline{\includegraphics[width=14cm]{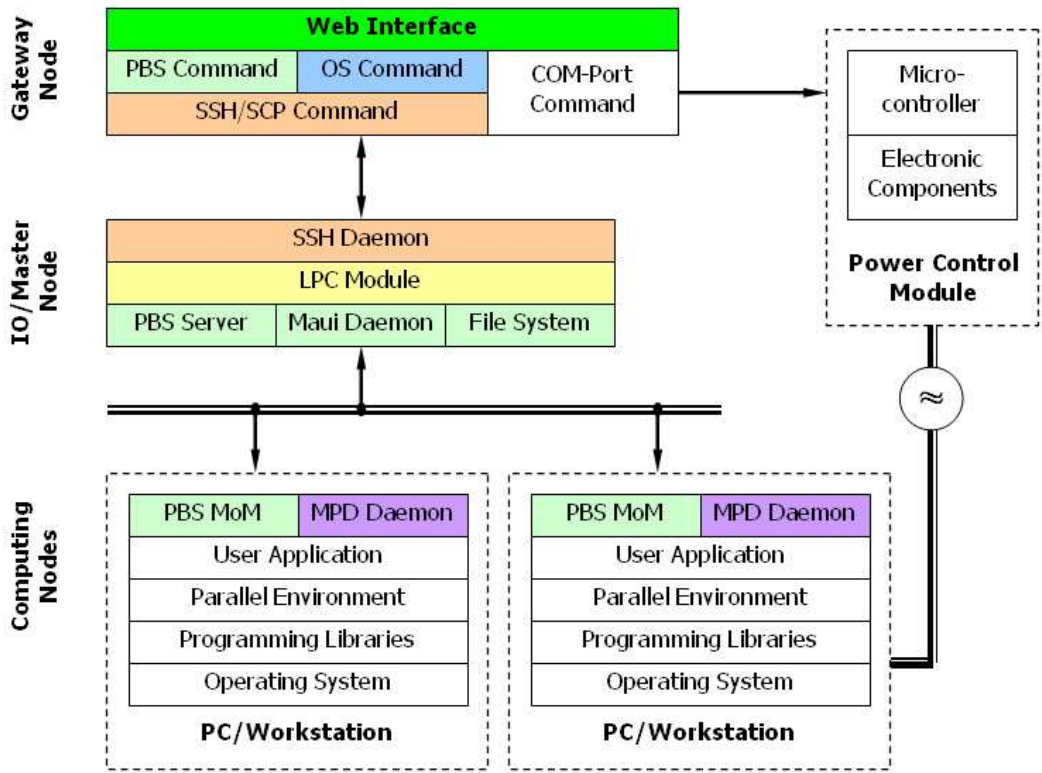}}
\caption{The openPC components and interactions implemented on LPC.}
\label{fig:interkoneksi}
\end{figure*}

Now we are ready to discuss the implementation on the LPC. As shown in Fig.
\ref{fig:interkoneksi}, LPC has 3 components :
\begin{enumerate}
\item Gateway node :\\
This is the front-end interfacing the users with the public cluster over an
integrated web-interface.
\item I/O / Master node : \\
This node is responsible for all computational jobs in LPC. It executes the
\texttt{pbs\_server} and maui daemon to manage the whole cluster resources.
\item Nodes :\\
The \texttt{pbs\_mom} is executed in these nodes. The PBS MoM is responsible for
communication with \texttt{pbs\_server}, that is providing places for jobs
submitted by \texttt{pbs\_server}, responding the resource monitoring requests,
reporting the finished jobs and so on.
\end{enumerate}

On the other hand, as mentioned before the web interface at LPC handles three
types of commands, the so called PBS, OS and com-port related commands. The
first one is responsible for the computational works being performed. The second
one is related to the file managements, while the last one is responsible for
hardware control and monitoring systems.

We have developed the so called LPC module as the public cluster module. Both
PBS and OS commands are sent to the computing nodes via this LPC module using
\texttt{ssh} protocol to ensure secure data transfer. We have deployed the RSA
or DSA key based passwordless \texttt{ssh}, rather than using the standard
Apache modules for the sake of security. In contrary, the com-port commands
directly trigger the external shell scripts to send certain signals to the
microcontrollers to control power modules etc. We have found that this mechanism
is much more reliable for our purposes than, for instance, real-time platform
like Java which could exhaust more resources.

Now let us discuss in more detail the LPC module, the implementation of the
block approach to build independent subsets of nodes, and the whole procedures
deployed in LPC.

\subsection{LPC module : public cluster module}

One of the main differences with another cluster / grid architectures is, as can
be seen in Fig. \ref{fig:interkoneksi}, there is an additional layer called as
LPC module in the I/O / master node. The main purposes of deploying this module
are :
\begin{itemize}
\item Load balancing : \\
Taking over all distributed computing from the gateway node. For instance,
although the command to access PBS is submitted by gateway node, the query to
the PBS is done by the LPC module.
\item Security :\\
Isolating the computing nodes completely from the gateway node. Since the
end-users access the system over a web interface installed in the gateway node,
this would keep them away from the computing nodes itself.
\item Easy maintenance :\\
Since the web interface and the distributed computing tools are separated at
different nodes, maintenance works at one of them would not influence the
others.
\end{itemize}

These purposes can actually be accomplished using the Remote Procedure Call
(RPC) model. In LPC it is realized using \texttt{ssh} that has been found more
secure than the others. Without this module, the architecture is similar to the
existing web interface as HPC2N which is also interfacing the PBS over web
\cite{hpc2n}. In contrary, using the LPC module, all (PBS and OS) commands
submitted from the gateway node are passed through this module. Thereafter, all
activities related to the file system and job management are also handled by the
module. Hence, the LPC module contains all pre-defined commands to recognize the
incoming commands and passed the authorized ones to the \texttt{pbs\_server},
maui daemon and file system. This additional layer improves the security since
the unauthorized commands are discarded immediately to prevent inappropriate
accesses by users.

\subsection{Defining a block}

Once a block of nodes has been activated, the host and user access lists
associated to the block are automatically created. The queue is exclusively
provided to a single user assigned to the block. This would ensure that the
queue will send the jobs only to the assigned block and nodes as discussed
before. This can be done using TORQUE and Maui by disabling the
\texttt{acl\_host\_enable} in the host access list. Further, the users who could
submit jobs to a particular queue are limited in the user access list.  \\

Below is a real example of the node and user mapping configuration in a queue
using Queue Manager (\texttt{Qmgr}) :
\begin{verbatim}
create queue block01 
set queue block01 queue_type = Execution 
set queue block01 acl_host_enable = False 
set queue block01 acl_hosts = node01 
set queue block01 acl_hosts += node04 
set queue block01 acl_hosts += node03 
set queue block01 acl_hosts += node02 
set queue block01 acl_user_enable = True 
set queue block01 acl_users = user01
set queue block01 resources_max.cput = 24:00:00 
set queue block01 enabled = True 
set queue block01 started = True 
\end{verbatim}
In this case, the queue name associated to a particular block is
\texttt{block01}. It is owned by a user \texttt{user01}, and consisting of 4
nodes : \texttt{node01}, \texttt{node02}, \texttt{node03} and \texttt{node04}.
This configuration realizes the requirements for LPC :
\begin{itemize}
\item Constraining \texttt{user01} to be able to access only 4 mentioned nodes
within a cluster \texttt{block01}. 
\item In contrary, another users are forbidden to access those nodes. That means
those nodes are accessible exclusively for the user \texttt{user01}.
\item The number of nodes and the nodes itself are fixed within the assigned
usage period.
\end{itemize}
Although the users can reboot their nodes, this kind of configurations is
automatically generated by the system. However, the users still have freedoms to
choose any parallel environments available in the LPC. Anyway, LPC at time being
provides standard MPI-based middlewares as OpenMPI \cite{openmpi}, LAM/MPI
\cite{lambmpi} and MPICH2 \cite{mpich2}.

\subsection{Procedures and job managements}

Here we describe the step by step procedures to perform a distributed computing
job on LPC. There are two main procedures : the first one is the block
activation by the administrator to turn up a block of nodes and make it
available for the approved user. Secondly, the job management by users to
perform any parallel jobs on the block allocated for them like compiling and
executing the parallel programming.

First of all, we list the procedures for a block activation :
\begin{itemize}
\item A registered user apply for a block with certain number of nodes, the
planned usage period and the detailed project description. We should note that
the registration procedure for new users is trivial such that no need to explain
it here. \\
\centerline{\includegraphics[width=7cm]{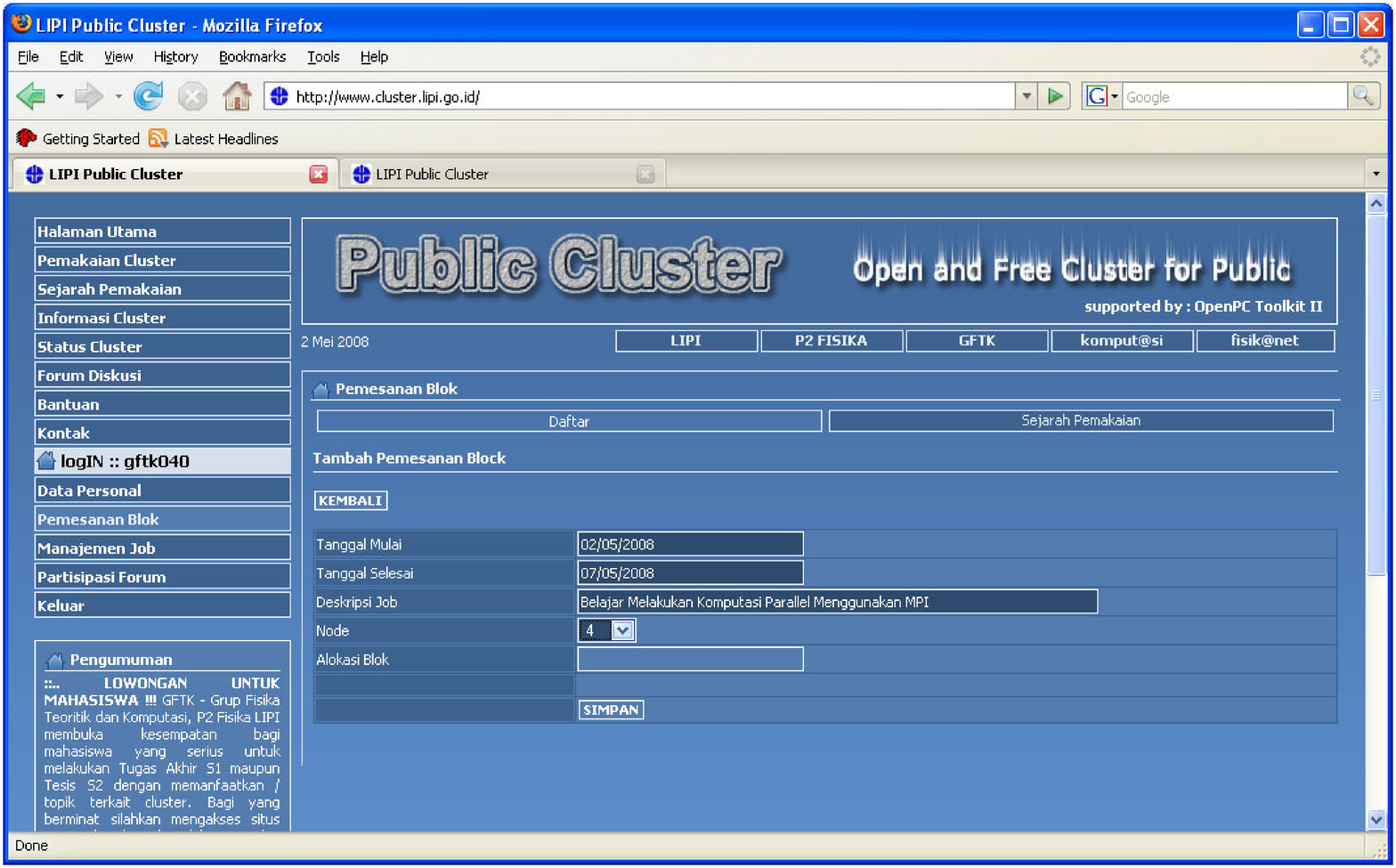}}
\item The administrator verifies each request. We always put an attention to the
congruity of the requested number of nodes with the planned job description. The
allocated nodes are thereafter the subject of availability at the time, and the
number is not necessary following the original request. \\
\centerline{\includegraphics[width=7cm]{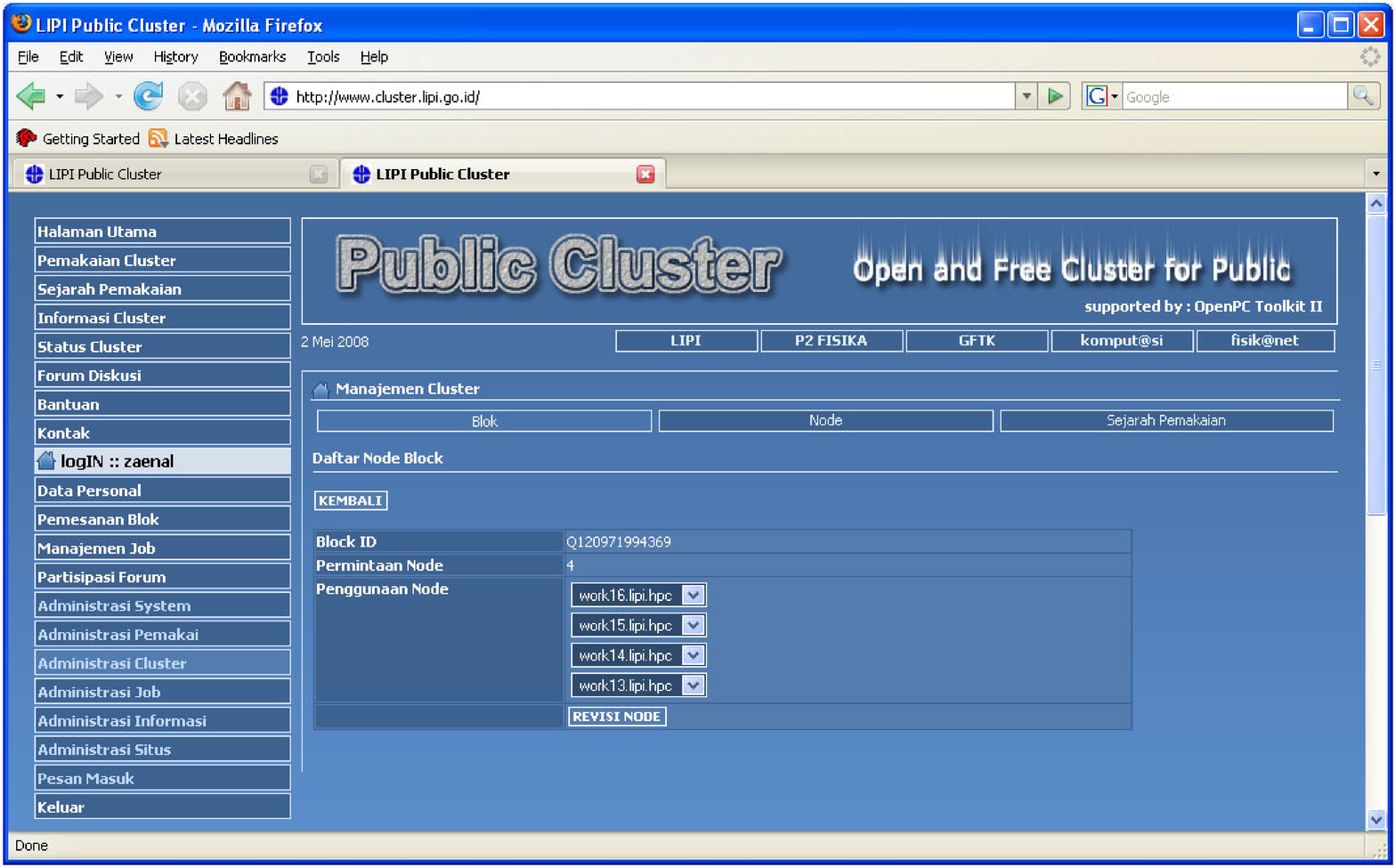}}
\item Once approved, the allocated block and nodes are activated by the
administrator. The nodes allocated for the block are automatically turned on,
and thereafter the system creates a particular queue by activating an access
list for certain nodes and user as the single owner of his / her block. \\
\centerline{\includegraphics[width=7cm]{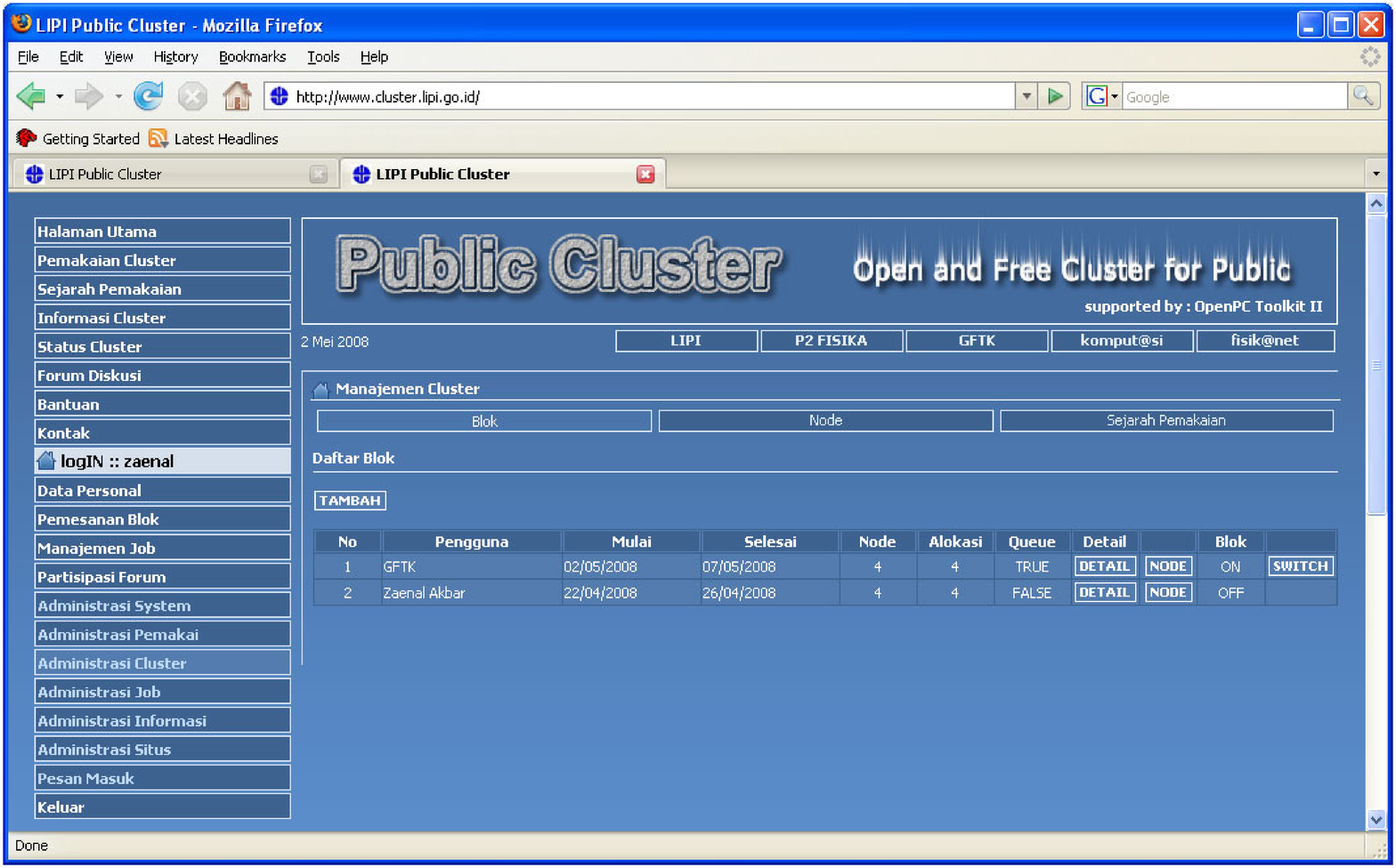}}
\item During the running period, the administrator have a full access to monitor
the block to ensure that everything works well. \\
\centerline{\includegraphics[width=7cm]{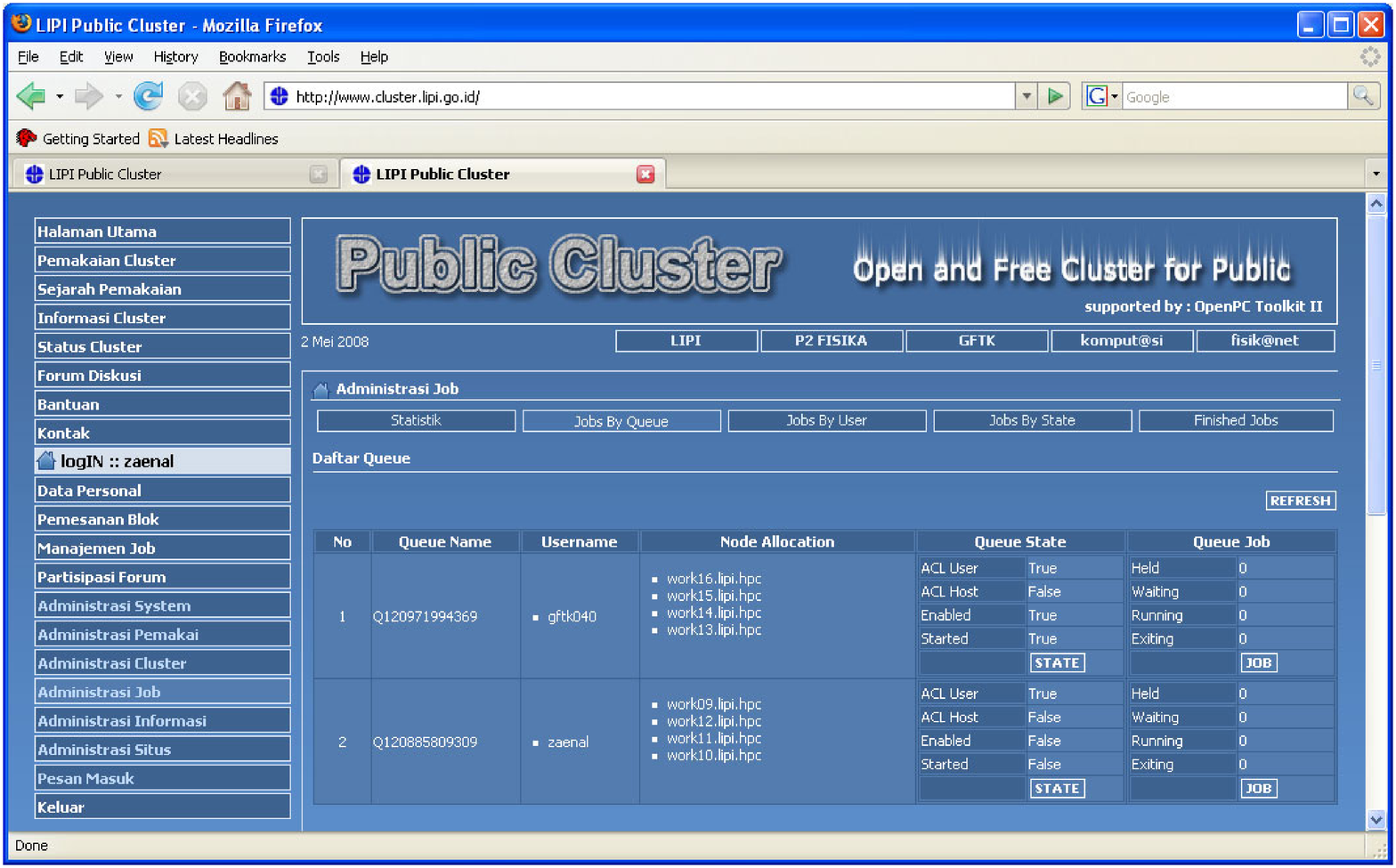}}
\end{itemize}

Once a user has been approved and assigned with a block of nodes, the user is
ready to perform any jobs on it. Below is the procedures of job management at
the LPC :
\begin{itemize}
\item The user should first choose a parallel environment from the available
ones at the LPC. This freedom is quite unique and enables users to perform
self-benchmarking of their own program on different parallel environments. \\
\centerline{\includegraphics[width=7cm]{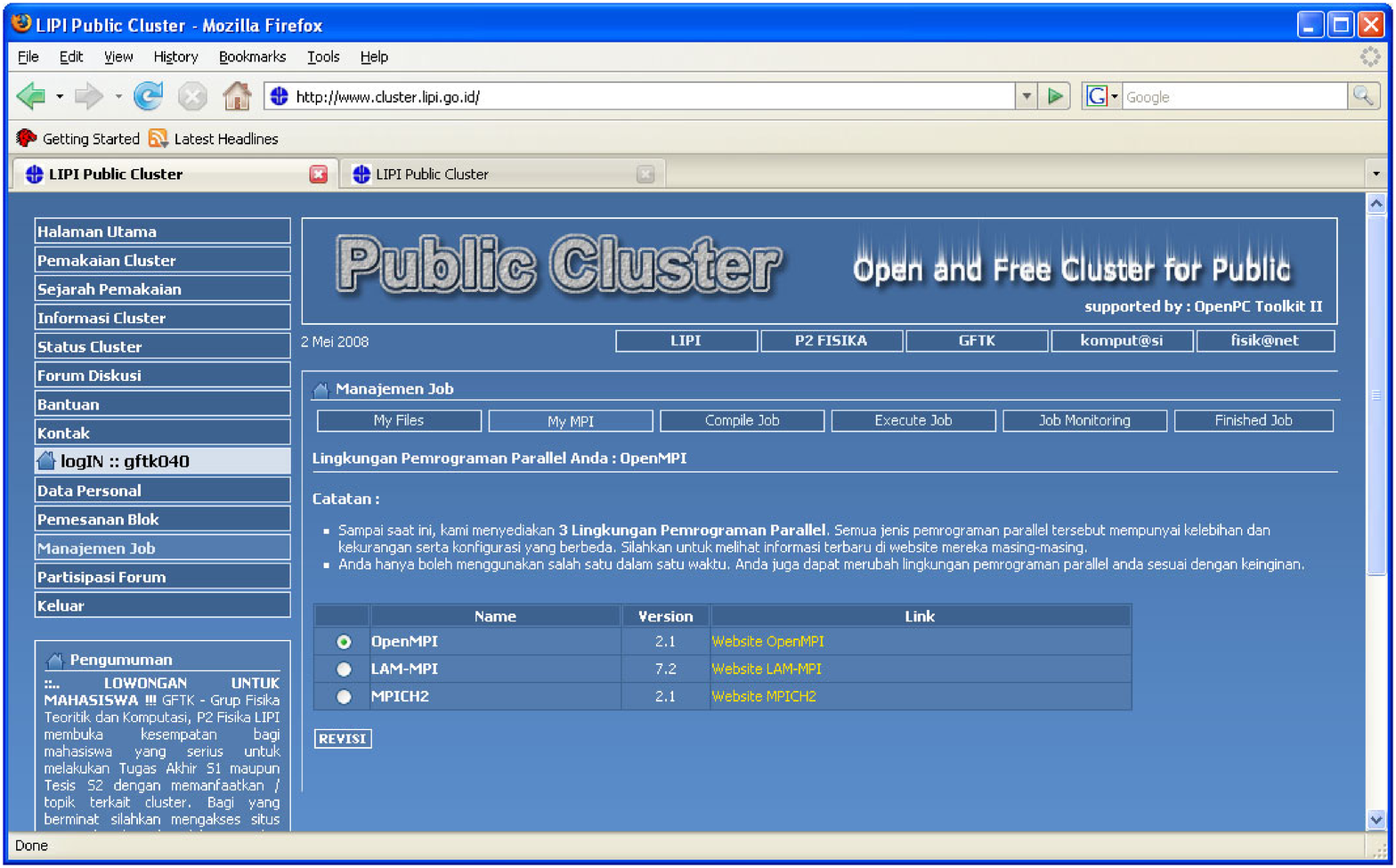}}
\item The program compilations and executions are done through resource manager,
in the case of LPC is TORQUE. Importantly, at this step the users are able to
put any desired options relevant for the parallel environment being used. Again,
the user is however forced to submit their jobs only within the allocated nodes
through the queue assigned for them. \\
\centerline{\includegraphics[width=7cm]{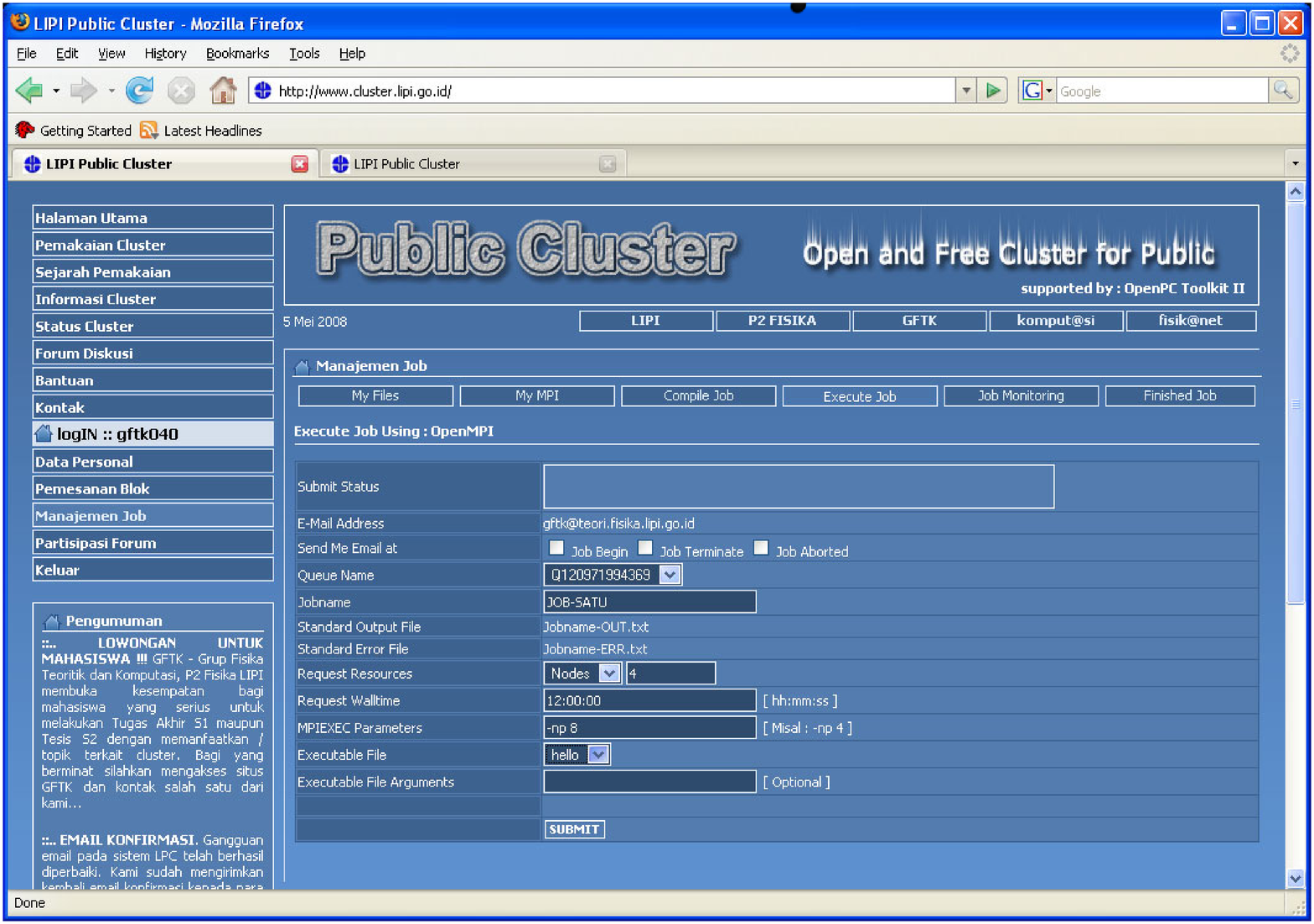}}
\item The user can monitor, re-execute, suspend, stop and delete their submitted
jobs. \\
\centerline{\includegraphics[width=7cm]{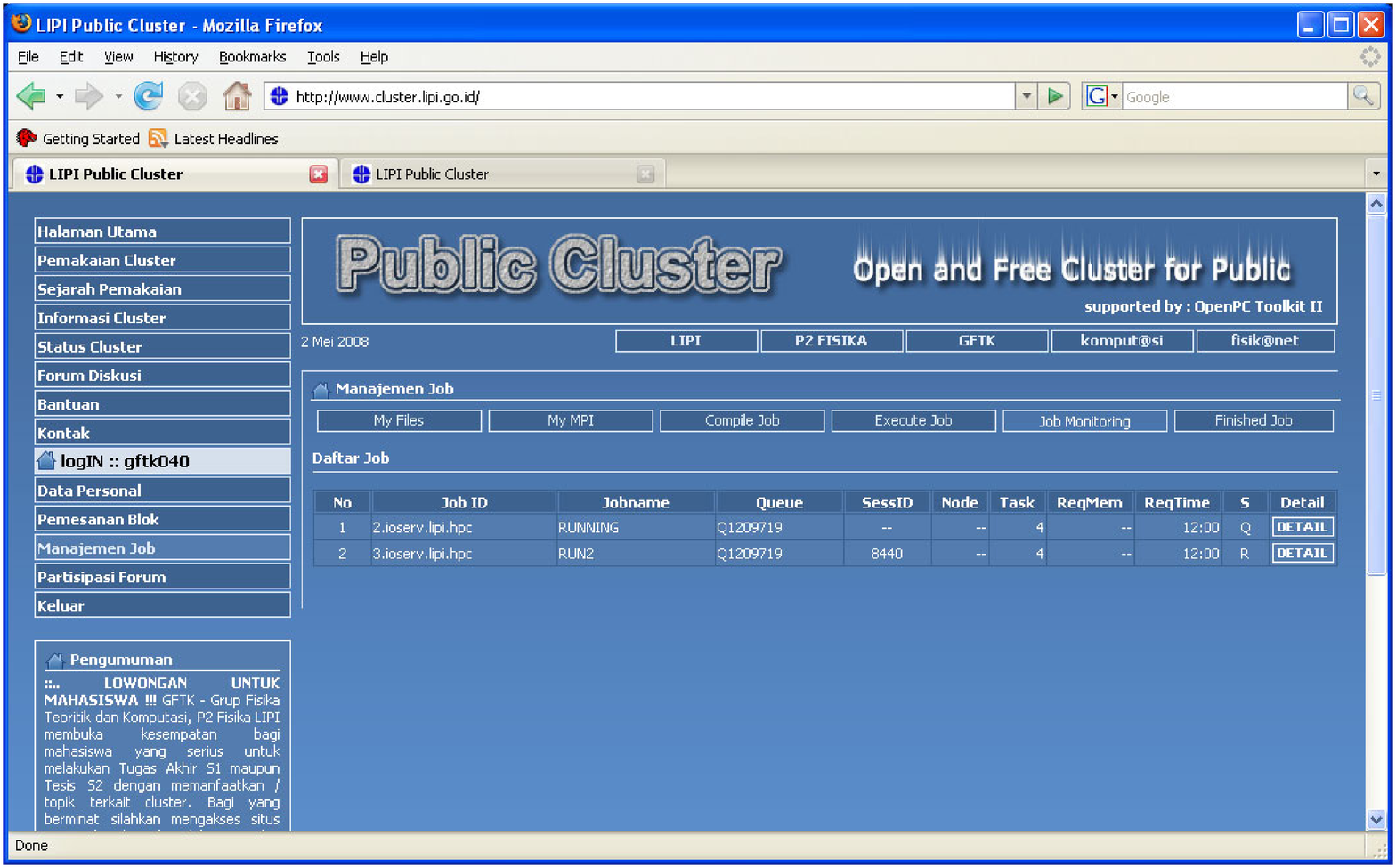}}
\item All executed jobs and its results including the detailed logs are stored
by the system for user access later on. The users can download and evaluate the
data subsequently. The detailed logs are generated as a job has been finished by
adding an additional command to the \texttt{epilog} script in each MoM daemon.
\\
\centerline{\includegraphics[width=7cm]{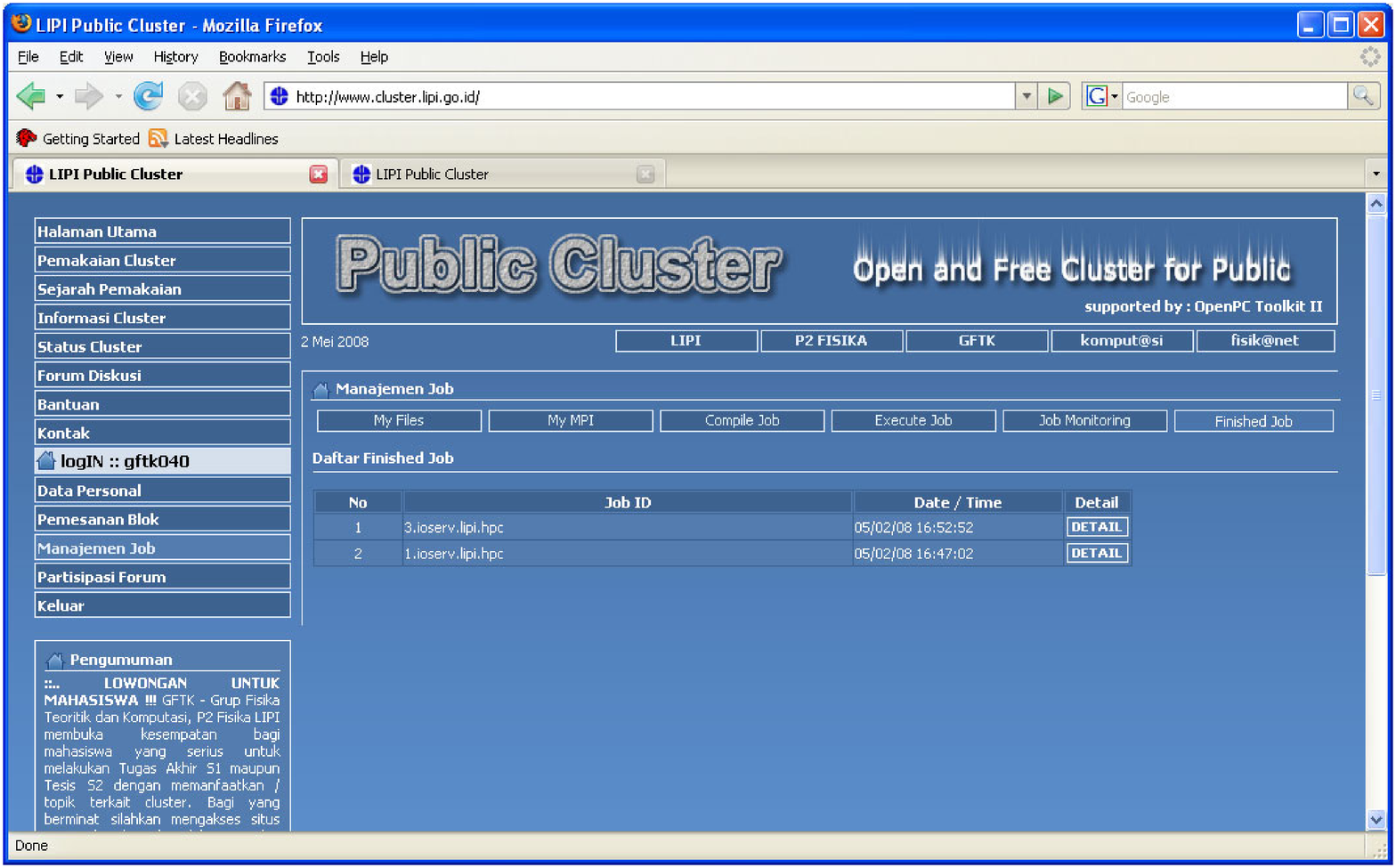}}
\end{itemize}

\begin{figure*}[t]
\centerline{\includegraphics[width=15cm]{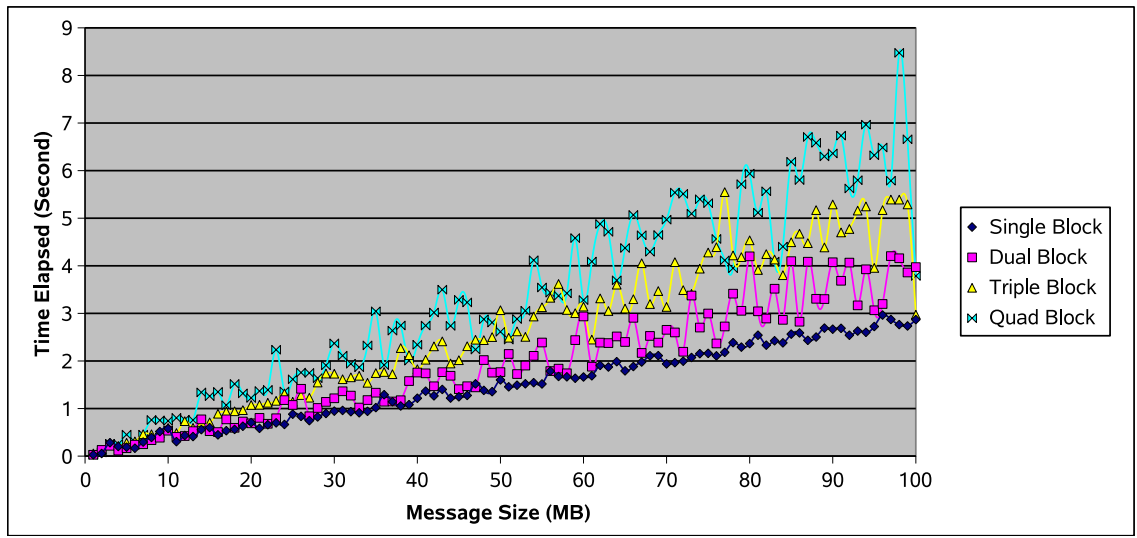}}
\caption{Performance analysis for multi-block approach with single master for
various block numbers.}
\label{fig:implementasi}
\end{figure*}

\section{PERFORMANCE}

So far, the LPC has enjoyed numerous visits from both local and international
visitors, especially from surrounding regions. Although at time being the
interface are available only in Bahasa Indonesia, we receive tenths new
registrations, and on an average ten percents are approved and becoming active
users over a month. Most of them are college students conform the initial
motivation of developing the LPC. Even not more than 10\% of total approved
users, some research groups also make use of LPC to perform much more advanced
computational works.

For the sake of completeness, let us present a simple performance analysis on
the reliability of multi-block approach before summarizing the paper. The
performance analysis is done for the case of multi-block approach using a single
master node as proposed originally in \cite{icts}. This approach is relevant for
small scale clusters with limited hardwares, and are in particular suitable for
processor / memory intensive computing works. So, it is quite interesting
mechanism for public clusters where the blocks are consisting of small number of
nodes. 

The benchmarking result is given in Fig. \ref{fig:implementasi}. It has been
done for $1 \sim 4$ block(s) of nodes with a single master node for all of them.
The benchmarking is done by flooding the data to all nodes simultaneously where
$30\%$ of data go through the master node. The data size is set to $1\sim 100$
MB with 1 MB incremental increase. The samples have been taken from the averages
of six measurements for each size of data. From the figure, obviously the
elapsed times are increasing as the flooded messages are greater. However for
moderate data size, namely less than $\sim 20$ MB that is the considerable level
of processor intensive works, the performance is still reliable. We can finally
argue that the multi-block approach with a single master node is a good
alternative for a small scale cluster divided into several independent blocks
with few nodes in each of them.

We should note that this approach has been deployed at LPC during the initial
period of implementation. Along with its development and various computational
works performed on it, we have later on deployed multi-block cluster with
independent master nodes in each block at LPC. Actually, we prefer to implement
both approaches at  LPC to optimize the existing infrastructures, since in some
cases the users needs only few nodes in a block for very small scale
computational works. The multi-block approach with a single master node suits
this level of needs. We have unfortunately not succeeded in implementing both
approaches simultaneously. Therefore, at time being all blocks at LPC have their
own master nodes, regardless with its number of nodes \cite{lpc}. There is another advantage on this architecture, that is it guarantees the total performance when the system is scaled up. This approach also improves the dependences among the blocks.

\section{CONCLUSION}

We have introduced the whole aspects of openPC, an open toolkit to enable public
access with full ownership to a cluster machine. We argue that such open
facility is important to broaden opportunities for, in particular, young people
and small research groups to perform and conduct advanced computing works using
HPC environments. According to the feedback from our users in Indonesia during
the last 3 years of implementation, the facility could provide a real experience
on HPC, either for educational or real scientific purposes, as if they had their
own HPC machines.

Now, we are entering the last phase of development. In this phase we are
developing the MPI-based middleware dedicated for this kind of cluster to
optimize the resources much  better using particular resource allocation and
management algorithms which fit its unique characteristics.

On the other hand, we also develop a web-service based module to connect a
particular block of nodes in a public cluster to a global grid with any types of
grid middlewares \cite{iccce}. Through this web service, the users of public
cluster are able to realize a collaboration with their global partners without
having their own HPC machines. This work is in progress.

Lastly, concerning global audiences, the English version of web-interface and
the complete manual is also under consideration. Since the whole system is now
open for public under GNU Public License (GPL),  volunteers to accomplish these
works are welcome through SourceForge.net \cite{openpc}.

\section*{Acknowledgment}

We greatly appreciate fruitful discussion with Slamet, B.I. Ajinagoro and G.J.
Ohara  throughout the initial period of the implementation of LPC. This work is
funded by the Riset Kompetitif LIPI in fiscal year $2007\sim2009$ under Contract
no. 11.04/SK/KPPI/II/$2007\sim2009$, and the Indonesian Toray Science Foundation
(ITSF) Research Grant 2006.

\bibliographystyle{IEEEtran}
\bibliography{klaster}

\begin{IEEEbiography}[{\includegraphics[width=1in,height=1.25in,clip,
keepaspectratio]{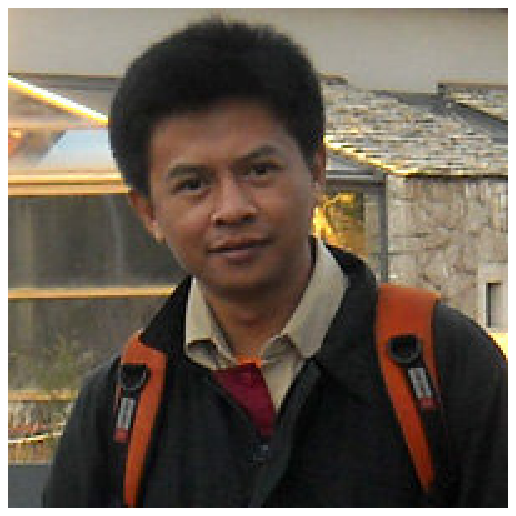}}]{ Z. Akbar}
 is a young researcher at the Group for Theoretical and Computational
Physics, Research Center for Physics, Indonesian Institute of Sciences (LIPI).
He is majoring on Computing Science, but his research is focused on the
distributing system. Since 2009 he also joined the Group for Bioinformatics and
Information Mining, Department of Computer and Information Science, University
of Konstanz as an associate researcher and working on distribution system for
bioinformatics problems. His URL is http://teori.fisika.lipi.go.id/~zaenal/.
\end{IEEEbiography}

\begin{IEEEbiography}[{\includegraphics[width=1in,height=1.25in,clip,
keepaspectratio]{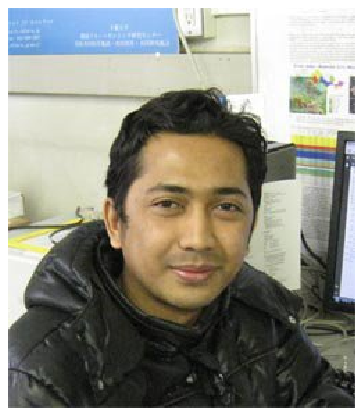}}]{I. Firmansyah}
 is a junior researcher at the Group for Theoretical and Computational
Physics, Research Center for Physics, Indonesian Institute of Sciences
(LIPI). Since 2009 he also joined the Center for Environmental Remote Sensing,
Chiba University as a research scientist. His research interest is mainly on
distributed instrumentation system. URL is
http://teori.fisika.lipi.go.id/~firmansyah/.
\end{IEEEbiography}

\begin{IEEEbiography}[{\includegraphics[width=1in,height=1.25in,clip,
keepaspectratio]{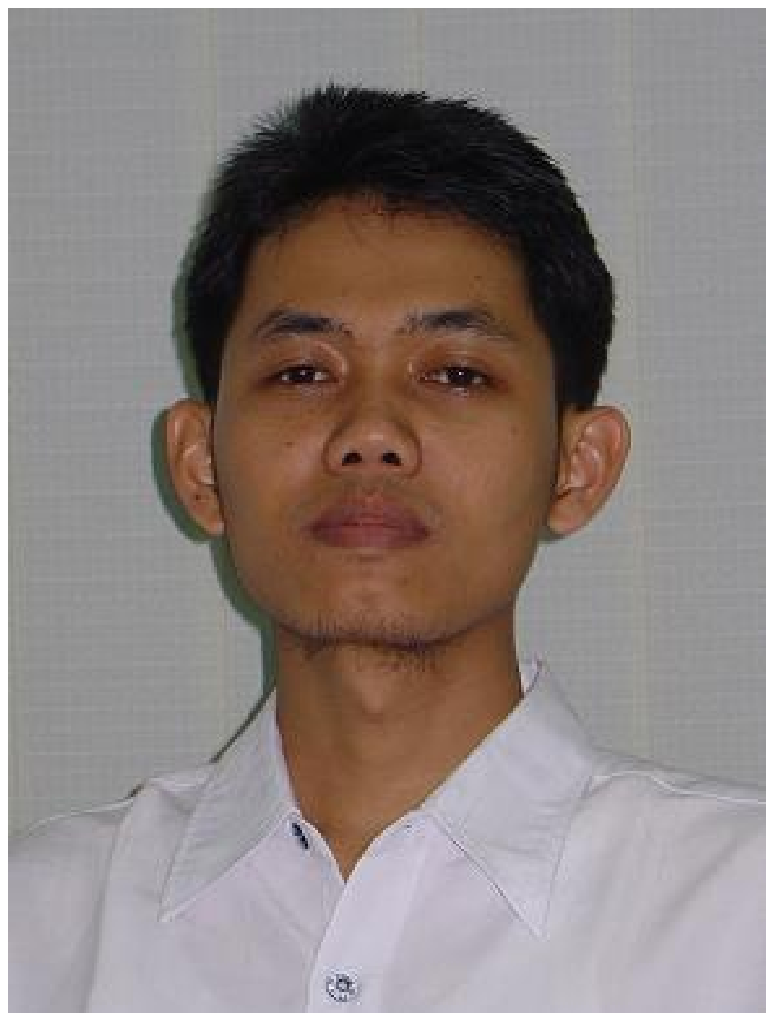}}]{B. Hermanto}
 is a junior researcher at the Group for Theoretical and Computational
Physics, Research Center for Physics, Indonesian Institute of Sciences (LIPI). 
His research interest is mainly on distributed networking system. 
His URL is http://teori.fisika.lipi.go.id/~hermanto/.
\end{IEEEbiography}

\begin{IEEEbiography}[{\includegraphics[width=1in,height=1.25in,clip,
keepaspectratio]{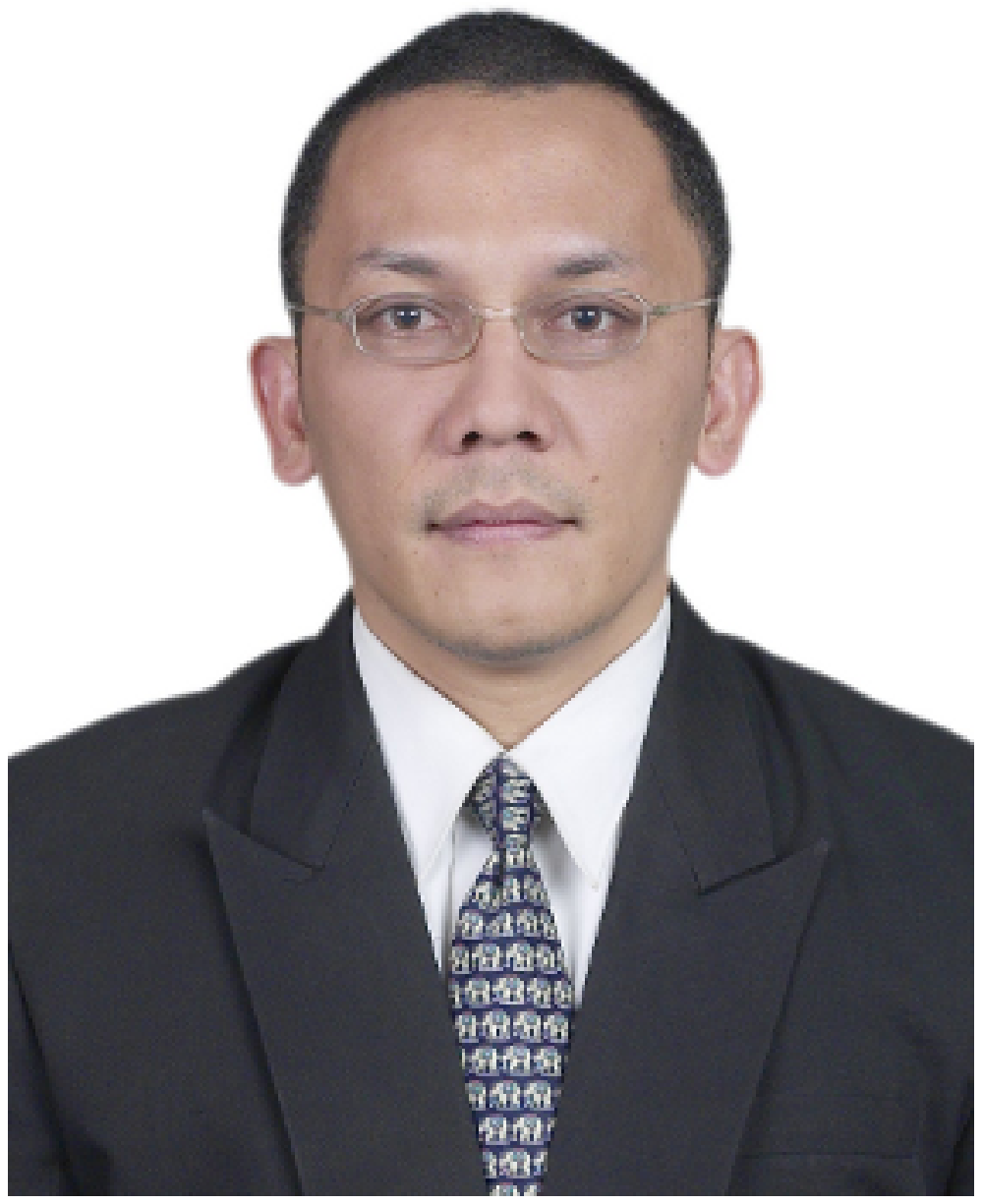}}]{L.T. Handoko}
 is a senior researcher and Group Head at the Group for Theoretical and
Computational Physics, Research Center for Physics, Indonesian Institute of
Sciences (LIPI). He is also a visiting professor at the Department of Physics,
University of Indonesia. His research interest covers broad area from
theoretical particle physics to computational science. His URL is
http://teori.fisika.lipi.go.id/~handoko/.
\end{IEEEbiography}

\end{document}